# POWER COMPARISON OF CMOS AND ADIABATIC FULL ADDER CIRCUITS

Y. Sunil Gavaskar Reddy[1] and V.V.G.S.Rajendra Prasad[2]

[1]Department of Electronics &Communication Engineering, Anurag Engineering College, JNTUniversity, Andhrapradesh,India
sunil_hanu123@ymail.com

[2]Department of Electronics &Communication Engineering, Anurag Engineering College, JNTUniversity, Andhrapradesh,India
veguntav@gmail.com

## ABSTRACT

*Full adders are important components in applications such as digital signal processors (DSP) architectures and microprocessors. Apart from the basic addition adders also used in performing useful operations such as subtraction, multiplication, division, address calculation, etc. In most of these systems the adder lies in the critical path that determines the overall performance of the system. In this paper conventional complementary metal oxide semiconductor (CMOS) and adiabatic adder circuits are analyzed in terms of power and transistor count using 0.18UM technology.*

## 1. INTRODUCTION

    Power minimization is one of the primary concerns in today VLSI design methodologies because of two main reasons one is the long battery operating life requirement of mobile and portable devices and second is due to increasing number of transistors on a single chip leads to high power dissipation and it can lead to reliability and IC packaging problems.

    The low-power requirements of present electronic systems have challenged the scientific research towards the study of technological, architectural and circuital solutions that allow a reduction of the energy dissipated by an electronic circuit. One of the main causes of energy dissipation in CMOS circuits is due to the charging and discharging of the node capacitances of the circuits, present both as a load and as parasitic. Such part of the total power dissipated by a circuit is called dynamic power. In order to reduce the dynamic power, an alternative approach to the traditional techniques of power consumption reduction, named adiabatic switching [1][2][14][18], has been proposed in the last years. In such approach, the process of charging and discharging the node capacitances is carried out in a way so that a small amount of energy is wasted and a recovery of the energy stored on the capacitors is achieved.

    In literature, various kinds of adiabatic circuits proposed [1][12][18][20] all of them can be grouped into two fundamental classes: fully adiabatic circuits and quasi-adiabatic or partial energy recovery circuits[1].In the first class, in particular working conditions can consume asymptotically zero energy for operation, the large area occupation and the design complexity make these circuits not competitive with traditional CMOS where as in second class circuits designed to recover large portion of the energy stored in the circuit node capacitances. This energy loss drawback however allows a good trade-off between circuit complexity and then area occupation.

Microprocessors and digital signal processors rely on the efficient implementation of generic arithmetic units to execute complicated algorithms like filtering, FFT. In these applications, multipliers are an important dissipation source. The basic element in multiplier is adder circuit so Therefore, power-efficient multipliers require power efficient implementation of adder circuit. In the literature different works presented on comparison of conventional CMOS adder circuits [3][5][15][16][17] and also comparison of adiabatic families[11][19][20][21]. In this work we analyzed the performance of conventional and adiabatic adder circuit's in-terms of power and transistor count.

The rest of the paper is organized as follows. Section 2 gives details of conventional charging and adiabatic charging principle, Section 3 explains different full adder implementations, section 4 simulation results and finally section 5 is conclusion.

## 2. ADIABATIC PRINCIPLE

The operation of adiabatic logic gate is divided into two distinct stages: one stage is used for logic evaluation; the other stage is used to reset the gate output logic value. Both the stages utilize adiabatic switching principle. In the following section conventional switching and adiabatic switching analyzed in detail.

### 2.1. Conventional Switching

There are three major sources of power dissipation in digital CMOS circuits those are dynamic, short circuit and leakage power dissipation. The dominant component is dynamic power dissipation and is due to charging, discharging of load capacitance [2]. The equivalent circuits of CMOS logic for charging and discharging is shown in Fig.1. The expression for total power dissipation is given by

$$P_{tot} = \alpha . C_L . V . V_{DD} . f_{clk} + I_{SC} . V_{DD} + I_{le} . V_{DD} \tag{1}$$

Equation (1), the first term represents the dynamic power, where $C_L$ is the loading capacitance, $f_{clk}$ is the clock frequency, and $\alpha$ is the switching activity. In most cases, the voltage swing V is the same as the supply voltage $V_{dd}$; however, in some logic circuits, the voltage swing on some internal nodes may be slightly less. The second term is due to the direct-path short circuit current $I_{sc}$ which arises when both the NMOS and PMOS transistors are simultaneously active, conducting current directly from supply to ground. Finally, leakage current $I_{le}$ which can arise from substrate injection and sub threshold effects is primarily determined by fabrication technology considerations. [2][4]

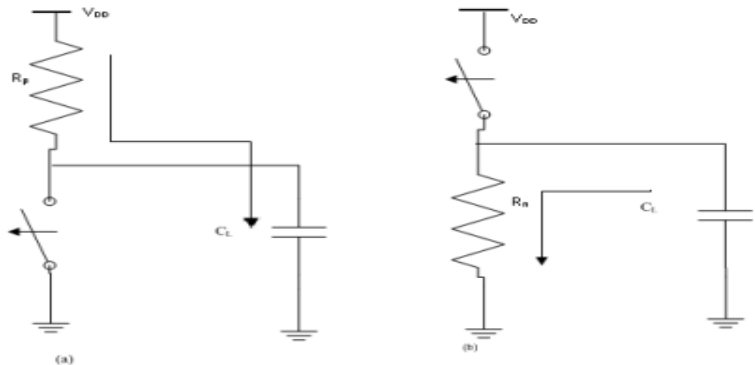

**Figure 1 Conventional CMOS a) Charging b) Discharging**

## 2.2. Adiabatic Switching

Adiabatic switching can be achieved by ensuring that the potential across the switching devices is kept arbitrarily small. This can be achieved by charging the capacitor from a time-varying voltage source or constant current source [1][4][9], as shown in Fig. 2. Here, R represents the on-resistance of the pMOS network. Also note that a constant charging current corresponds to a linear voltage ramp. Assuming that the capacitance voltage $V_C$ is zero initially, the variation of the voltage as a function of time can be found as

$$V_C(t) = I_S \cdot t / C \qquad (2)$$

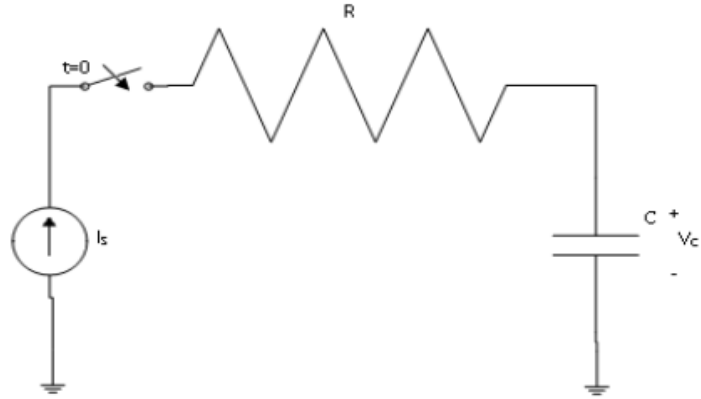

**Figure 2 Schematic for adiabatic charging process**

Hence the charging current can be expressed as a function of $V_C$ and time t

$$I_S = C \cdot V_C(t) / t \qquad (3)$$

The amount of energy dissipated in the resistor R from t = 0 to $t$ = T can be found as

$$E_{diss} = R \int_0^T I_S^2 dt = R I_S^2 T \qquad (4)$$

Combining (3) and (4), the dissipated energy during this charge-up transition can also be expressed as

$$E_{diss} = \frac{RC}{T} \cdot C V_C^2(T) \qquad (5)$$

From (5) we can say that the dissipated energy is smaller than for the conventional case if the charging time T >>2RC and can be made small by increasing the charging time. A portion of the energy thus stored in the capacitance can also be reclaimed by reversing the current source direction, allowing the charge to be transferred from the capacitance back into the supply. Adiabatic logic circuits thus require non-standard power supplies with time-varying voltage, also called pulsed power supplies. The additional hardware overhead associated with these specific power supply circuits is one of the design trade-off. Practical supplies can be constructed by using resonant inductor circuits. But the use of inductors should be limited from

integrated circuit point because of so many factors like chip integration, accuracy, efficiency etc. [4]

An alternative to using pure voltage ramps is to use stepwise supply voltage waveforms, where the output voltage of the power supply is increased and decreased in small increments during charging and discharging. Since the energy dissipation depends on the average voltage drop across the resistor by using smaller voltage steps the dissipation can be reduced considerably [4]. The total dissipation using step wise charging is given by (6)

$$E_{tdiss} = \frac{1}{n} C V_{DD}^2 / 2 \qquad (6)$$

Where n is number of steps used to charge up capacitance to $V_{DD}$.

In literature, adiabatic logic circuits classified into two types: full adiabatic and quasi or partial adiabatic circuits. Full-adiabatic circuits have no non-adiabatic loss, but they are much more complex than quasi-adiabatic circuits. Quasi-adiabatic circuits have simple architecture and power clock system. There are two types of energy loss in quasi-adiabatic circuits, adiabatic loss and nonadiabatic loss. The adiabatic loss occurs when current flows through non-ideal switch, which is proportional to the frequency of the power-clock. If any voltage difference between the two terminals of a switch exists when it is turned on, non-adiabatic loss occurs. The non-adiabatic loss, which is independent of the frequency of the power-clock, is proportional to the node capacitance and the square of the voltage difference. Several quasi-adiabatic logic architectures have been reported, such as ECRL, 2N-2N-2P, PFAL etc.[1][9][12][13][14]

## 3. ADDER IMPLEMENTATION

A basic cell in digital computing systems is the 1-bit full adder which has three 1-bit inputs (A, B, and C) and two 1-bit outputs (*sum* and *carry*). The relations between the inputs and the outputs are expressed as

$$Sum = \overline{A}\overline{B}C + \overline{A}B\overline{C} + A\overline{B}\overline{C} + ABC \qquad (7a)$$

$$Carry = AB + BC + CA \qquad (7b)$$

### 3.1 Conventional adder [2], [4], [6]

Conventional CMOS Implementation consists of two functional blocks pull-up and pull-down. Pull-up functional block is implemented with P-channel MOS transistors and pull-down functional block is implemented with N-channel MOS transistors. In order to get symmetrical structure (7a) is rearranged as (8) and sum and carry implementation is shown in Fig.3.

$$Sum = ABC + (A + B + C)\overline{Carry} \qquad (8)$$

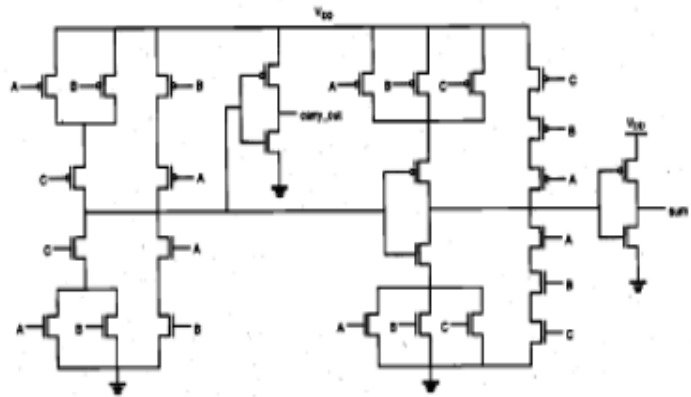

**Figure 3 Conventional Symmetrical CMOS Full Adder**

### 3.2 Pass transistor logic (PL) based adder [6]

Pass transistor logic is one of the well known nMOS logic style used to implement different functions. General method for deriving pass transistor logic diagram for a function is choosing control variable and pass variable based on the functional description. The main concept behind Complementary PL (CPL) is the use of only an nMOSFET network for the implementation of logic functions. This results in low input capacitance and high speed operation. The schematic diagram of the CPL full adder circuit is shown in Fig 4. Because the high voltage level of the pass-transistor outputs is lower than the supply voltage level by the threshold voltage of the pass transistors, the signals have to be amplified by using CMOS inverters at the outputs. CPL circuits consume less power than conventional static circuits because the logic swing of the pass transistor outputs is smaller than the supply voltage level.

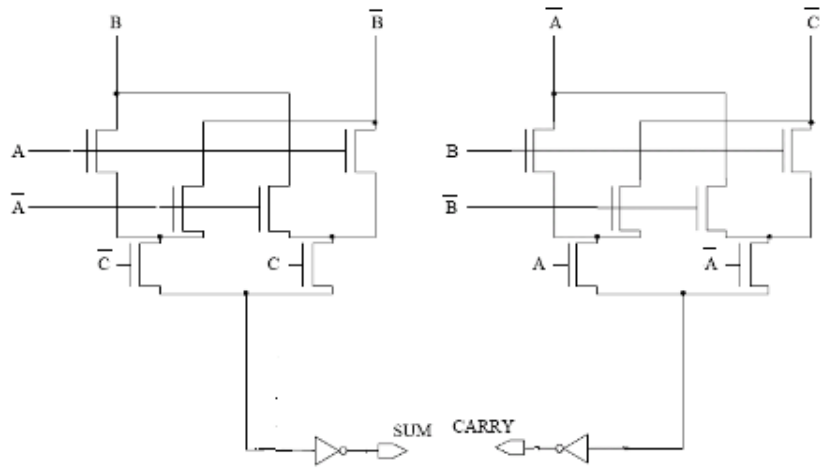

**Figure 4 Pass transistor based full adder**

### 3.3 Transmission gate (TG) based adder [4]

Transmission gate approach is another widely used CMOS design style to implement digital function. Transmission gate based implementation is similar to pass transistor with the difference that transmission gate logic uses nMOS and pMOS transistors where as pass transistor logic uses only one type of transistor i.e. either nMOS or pMOS. Full adder implementation based on TG logic is shown in Fig.5. [4]

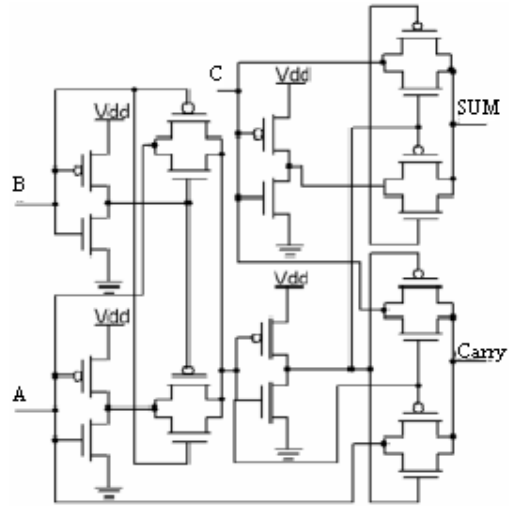

**Figure 5 Transmission gate based full adder**

## 3.4 10T adder or Static Energy Recovery Full adder (SERF)[8][15][19]

In the 10T adder cell, the implementation of XOR and XNOR of A and B is done using pass transistor logic and an inverter is to complement the input signal A. This implementation results in faster XOR and XNOR outputs and also ensures that there is a balance of delays at the output of these gates. This leads to less spurious SUM and Carry signals. The capacitance at the outputs of XOR and XNOR gates is also reduced as they are not loaded with inverter. If the signal degradation at the SUM and Carry is significant for deep sub-micron circuits, drivers can be used to reduce the degradation. The driver will help in generating outputs with equal rise and fall times. This results in better performance regarding speed, low power dissipation and driving capabilities. The output voltage swing will be equal to the $V_{DD}$, if a driver is used at the output. The circuit diagram of static energy recovery adder is shown in fig 6

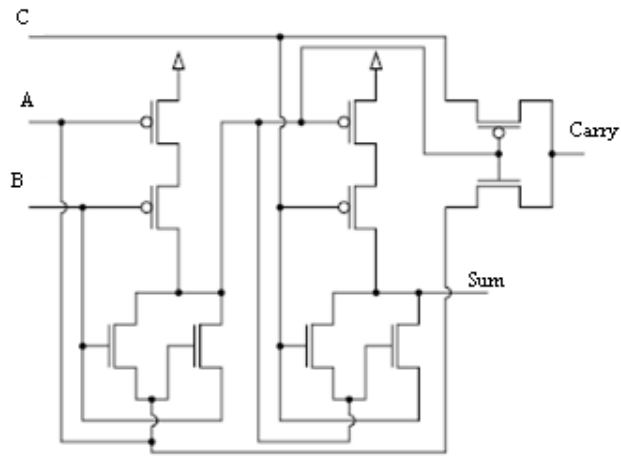

**Figure 6:Static energy recovery adder**

## 3.5 Pass transistor based adiabatic adder [7]

The sum and carry expressions for one bit full adder is given by

$Sum = A \oplus B \oplus C$

$Carry = AB + BC + CA$ (9)

The above equations can be re arranged as

$Sum = (A \oplus B)\overline{C} + \overline{(A \oplus B)}C$

$Carry = (A \oplus B)C + \overline{(A \oplus B)}B$ (10)

The sum and carry expressions in (10) have common terms and can be implemented using Fig.7. [7]

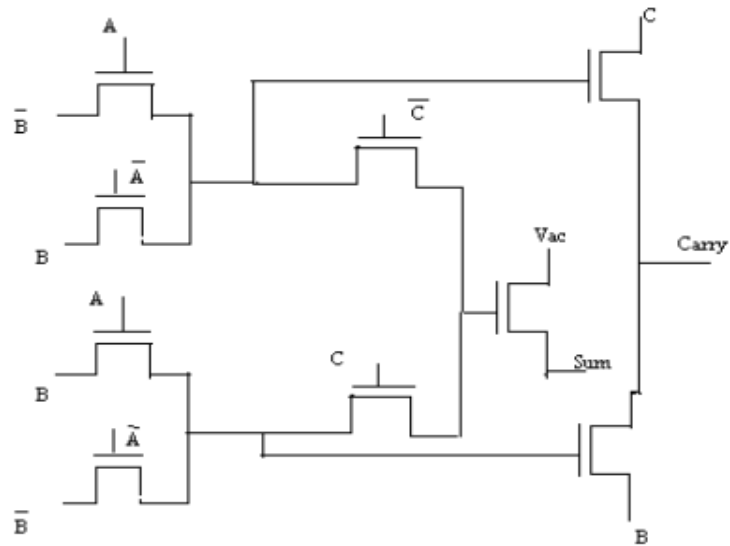

**Figure 7 Pass transistor based adiabatic full adder**

## 3.6 Positive feedback adiabatic logic (PFAL) adder [12][18][21]

The general PFAL gate consists of a two cross coupled inverters and two functional blocks F and /F (complement of F) driven by normal and complemented inputs which realizes both normal and complemented outputs. Both the functional blocks implemented with n channel MOS transistors. The equations used to implement PFAL adder given by (11) and the corresponding sum and carry implementations are shown in Fig.8a and Fig.8b

$Sum = \overline{A}\overline{B}C + \overline{A}B\overline{C} + A\overline{B}\overline{C} + ABC$

$Carry = AB + BC + CA$ (11)

## 3.7 Transmission gate based adiabatic adder (TGAL)

The general block diagram of transmission gate based adiabatic logic consists of two functional blocks F and complement of F operated with single clock power supply. Both normal and complemented inputs are available to functional blocks. Functional blocks are implemented using transmission gate or pass gate. The sum and carry implementation using transmission gate logic is shown in Fig .9a and Fig.9b. [4][11]

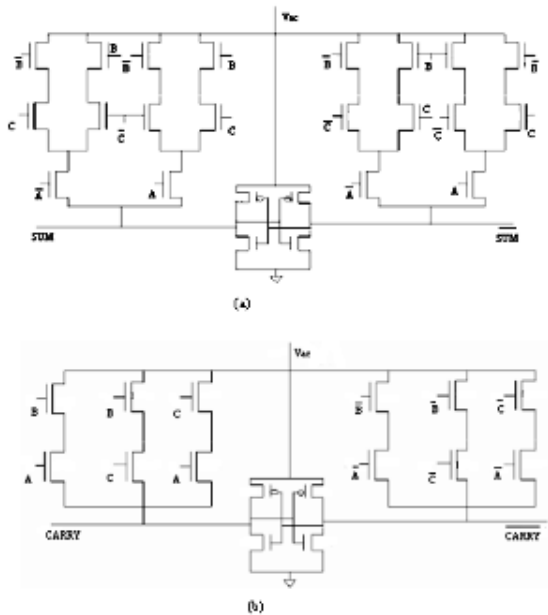

**Figure 8 Positive feedback adiabatic full adder a) Sum b) carry**

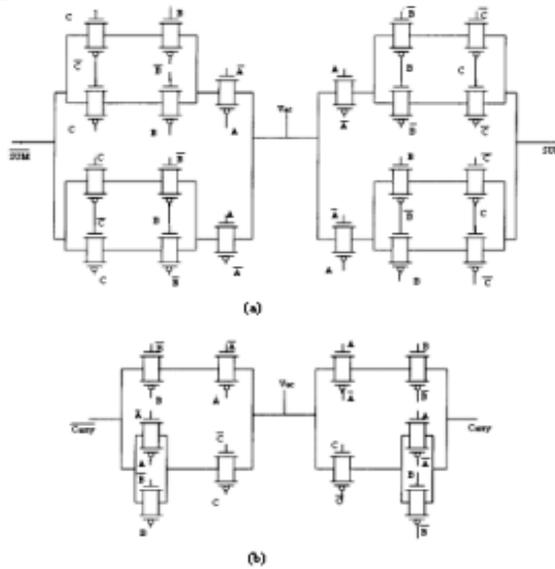

**Figure 9 Transmission gate based adiabatic full adder a) Sum b) Carry**

## 4 SIMULATION RESULTS

In order to estimate the power dissipation of the different circuits present in previous section we used power meter simulation model present in [4] and RC model present in [10].The schematic diagram of the test setup to estimate the power dissipation is shown in fig.10 and it contains DUT (design under test), linear current controlled current source, resistor, capacitor connected in parallel. All the circuits are designed and simulated using Tanner tools(S-edit, TSPICE) with 0.18um technology parameters.Fig.11 shows the simulation results of comparison of transmission gate adiabatic logic and static energy recovery full adder circuits. Table. I gives power dissipation values under the operating conditions $V_{DD}$=1.8V, $C_L$=20fF and frequency 50MHz.Fig.12 shows the variation of power dissipation with the frequency for the different circuit implementations.

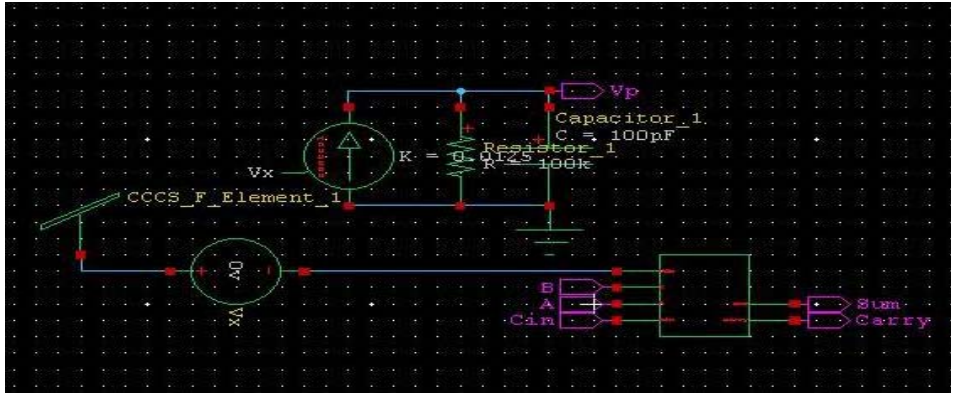

**Figure 10 power meter simulation setup for conventional CMOS adder**

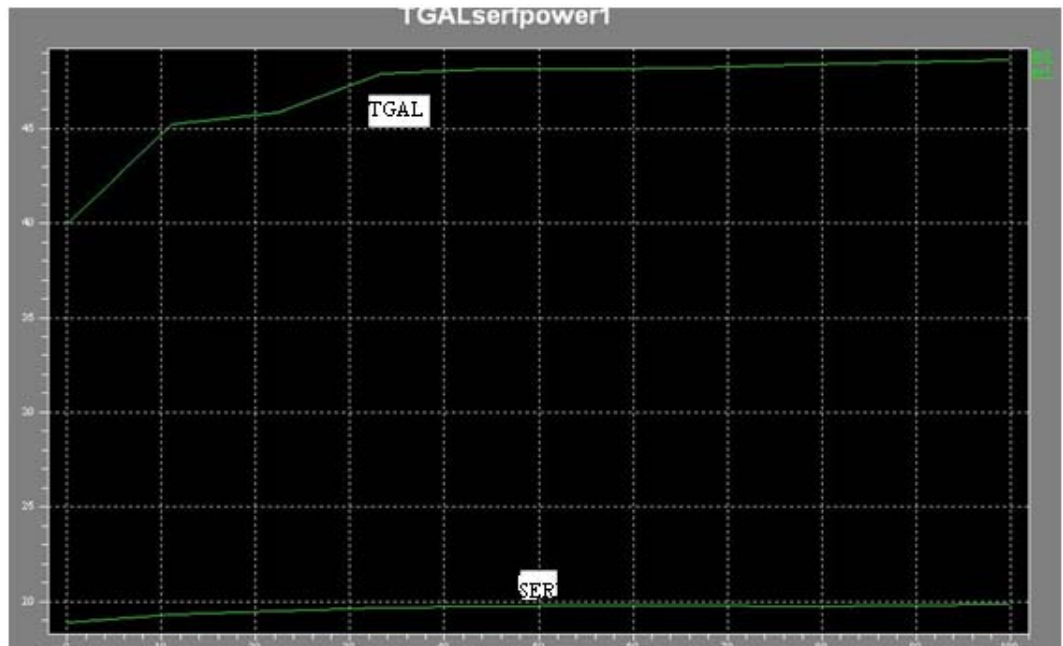

**Figure 11Comparision of TGAL and SERF power with frequency variation from o to 100MHz**

**Table 1 Power dissipation of various adders under the operating conditions ($V_{DD}$=1.8v, $C_L$=20fF, $f_{clk}$= 50MHz)**

| Parameter | ADDER TYPE | | | | | | |
|---|---|---|---|---|---|---|---|
| | CMOS | PL | TGL | PAL | PFAL | TGAL | SERF |
| Transistor Count | 28 | 22(24[a]) | 20 | 10 | 38 | 60 | 10 |
| Power dissipation(µW) | 1.9 | 1.2 | 2.1 | 0.06 | 0.05 | 0.85 | 0.08 |

[a] if optional pMOS transistors are considered

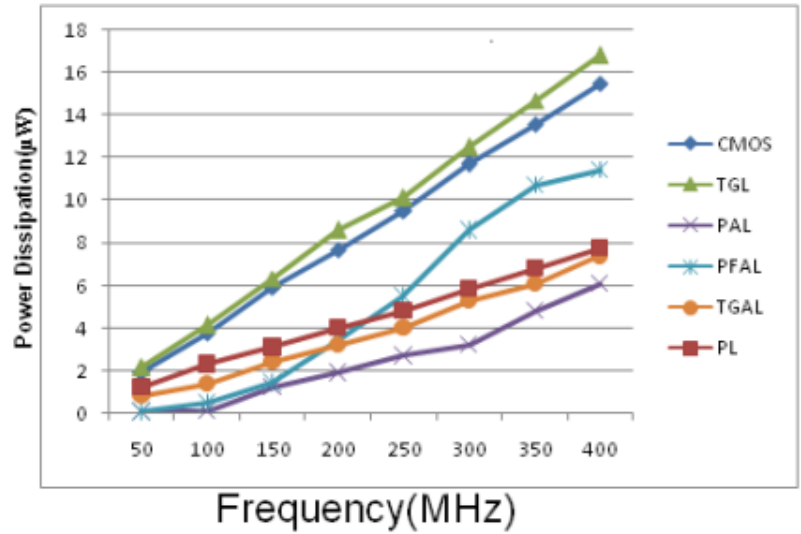

**Figure 12 Variation of power dissipation with frequency for different full adder circuits**

## 5. CONCLUSION

In this paper we compared the performance of different adiabatic logic adder circuits with traditional CMOS adder circuits. The analysis shows that designs based on adiabatic principle gives superior performance when compared to traditional approaches in terms of power even though their transistor count is high in some circuits so for low power and ultra low power requirements adiabatic logic is an effective alternative for traditional CMOS logic circuit design.

**Authors**

V.V.G.S.Rajendra Prasad completed his B.Tech from Andhra University and M.Tech from JNTU,Hyderabad with specialization VLSI System Design

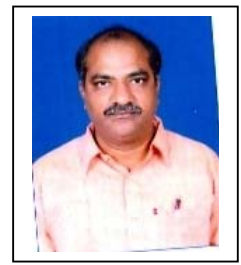

Y.Sunil Gavaskar Reddy Completed his B.Tech from JNTU Hyderabad and M.S from, IIIT Pune with specialization Advance microelectronics & VLSI Design.

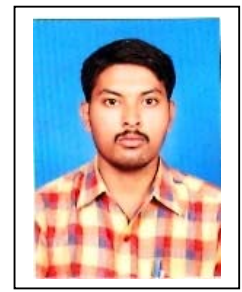